\documentclass[aps,twocolumn]{revtex4}%
\usepackage{amsfonts}
\usepackage{amsmath}
\usepackage{graphicx}
\usepackage{amssymb}%
\begin{document}
\title{Dispersive measurements of superconducting qubit coherence with a fast,
latching readout}
\author{I. Siddiqi, R. Vijay, M. Metcalfe, E. Boaknin, L. Frunzio, R.J. Schoelkopf, and M.H. Devoret}
\affiliation{Departments of Applied Physics and Physics, Yale University, New Haven,
Connecticut 06520-8284}

\begin{abstract}
The \textquotedblleft quantronium\textquotedblright \space is a
superconducting qubit consisting of a split Cooper pair box in which
a large tunnel junction is inserted. This circuit has a special bias
point where the Larmor frequency is, to first order, insensitive to
fluctuations in the bias parameters -- the charge of the box island
and the phase of the large junction. At this optimal working point,
the state of the qubit can be determined by dispersive measurements
that probe the second derivative of the state energy with respect to
these bias parameters. We use the quantronium phase degree of
freedom to perform a non-linear, dispersive measurement of its
inductive response using bifurcation amplification. This novel
readout projects the state of the qubit in a few nanoseconds, and
its latching property allows us to record the resulting information
in a few hundred nanoseconds. We have measured, using this
technique, Rabi oscillations and Ramsey fringes with an improved
signal to noise ratio and contrast. The speed of this new readout
scheme also opens the door for a new class of experiments which
would characterize the relaxation processes associated with the
measurement protocol.

\end{abstract}
\maketitle

\section{Introduction}

\bigskip

Superconducting tunnel junction circuits were first proposed for
quantum information processing several years ago, and at present,
are the most advanced solid state qubits with the longest measured
coherence times
\cite{quantronium,Delft-recent-echo,Semba,Clarke,Rob-recent,Martinis-recent,Nakamura-recent}%
. Yet, the physical origin of the noise sources limiting coherence
are still debated, even though the theoretical formalism for
treating the effects of noise in general is well developed
\cite{vanHarlingen,Shnirman,Esteve-theory}. It has been conjectured
that impurities or defects found on-chip could act as such noise
sources \cite{Martinis-resonances}. These parasitic elements may
exist in the junction tunnel barriers, the metallic electrodes, the
circuit substrate, or in some combination thereof. In addition, the
shadow-mask evaporation technique used to fabricate many
superconducting qubits typically generates extra electrodynamic
resonators in close proximity to the qubit junctions \cite{Esteve}.
These resonators can have a characteristic frequency comparable to
the qubit Larmor frequency, and are thus suspected to decohere the
qubit. The precise manner in which a qubit interacts with
uncontrolled degrees of freedom in its environment depends on the
topology of the tunnel junction circuit and how information is
written to and read from the qubit. Circuits which have a high
degree of symmetry can be significantly decoupled from a noisy
environment \cite{Ioffe,quantronium} when biased at special
operating points. The choice of readout scheme is also highly
significant. Dispersive measurements of the qubit state
\cite{Lupascu,Rob} probe the reactive part of the response of the
circuit, and are thus attractive since they minimally excite the
spurious degrees of freedom described above.

We report the first coherence measurements of a superconducting
qubit with a non-linear dispersive readout. Our approach involves
coupling the \textquotedblleft quantronium\textquotedblright\ qubit
\cite{quantronium} to a novel superconducting amplifier -- the
Josephson bifurcation amplifier (JBA) \cite{Siddiqi-JBA-PRL}. The
JBA is based on a non-linear electrodynamic resonator with two
metastable oscillation states \cite{Dykman}. In order to perform a
readout, the resonator is RF-energized to a level where its
oscillation state now acts as a sensitive pointer of the qubit
state. This technique does not generate any dissipation on chip
since the
resonator is only damped by circuitry outside the chip, i.e. a 50\thinspace$\mathrm{\Omega}%
$\ transmission line with a matched circulator and amplifier, and
enables a high-fidelity qubit readout with a MHz repetition rate. We
have measured Rabi oscillations and Ramsey fringes with sufficient
speed that real time filtering to correct for drifts in the charge
and flux bias becomes possible. Also, several successive readouts
may be performed within the energy relaxation time of the qubit
$(T_{1})$. This gives valuable information on the readout-induced
interaction between the qubit and its environment, and accounts for
the observed contrast.

\bigskip

\begin{figure}[t]
\includegraphics[width=3.1in]{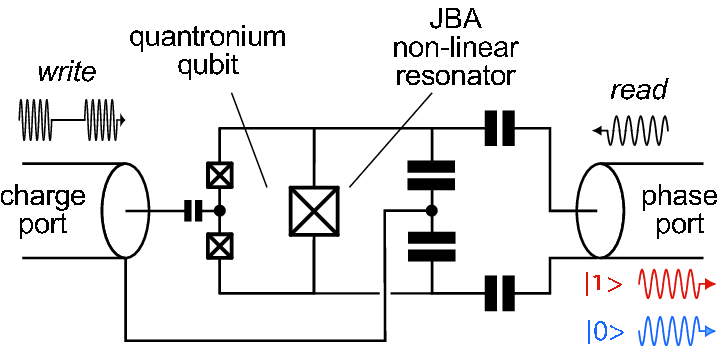}\caption{Schematic of the measurement
setup. The quantronium qubit is a split Cooper pair box with two
small Josephson junctions in which a large junction is inserted for
readout. This latter junction is shunted by two capacitors in series
and forms the non-linear resonator of the JBA readout. The qubit
state is manipulated by sending pulses to the charge port, while
readout operation is performed by sending a pulse to
the phase port and analyzing the phase of reflected signal, which carries information on the
qubit state.}%
\label{FigSampleWL}%
\end{figure}

\section{The hamiltonian of the quantronium qubit with JBA readout}

The principle of our experiment is schematically depicted in Fig. 1
and is based, as discussed above, on the quantronium qubit, a three
junction circuit which is analogous to a one-dimensional atom. The
set of three junctions consists of two small junctions, which we
assume to be identical and have a Josephson energy comparable to the
charging energy of the island between them, and a large junction,
whose Josephson energy is approximately 100 times larger than that
of each small junction. The gauge-invariant phase difference
$\hat{\theta}$ of the island with respect to the mid-point of the
capacitance shunting the large junction is analogous to the position
of the electron relative to the nucleus of the atom, while the
gauge-invariant phase difference $\hat{\delta}$ across the large
junction is the absolute position of the nucleus. Neglecting the
dissipation induced in the transmission lines, the total hamiltonian
of the split Cooper pair box with a JBA resonator is
$\hat{H}(t)=\hat{H}_{box}(t)+\hat{H}_{res}(t)$ with
\begin{widetext}
\begin{equation*}
\hat{H}_{box}(t)=4E_{C}\left(
\hat{N}-\frac{1}{2}+\frac{C_{g}U(t)}{2e}\right) ^{2}-\left(
E_{J}\cos \frac{\hat{\delta}}{2}\right) \cos \hat{\theta}\qquad
;\qquad \hat{H}_{res}(t)=\frac{\hat{Q}^{2}}{2C}-E_{J}^{R}\cos
\hat{\delta}-\varphi_{0}I(t)\hat{\delta}
\end{equation*}
\end{widetext}
 Here, $\hat{N}$ and $\hat{Q}/2e$ are the momenta conjugate to the
generalized positions $\hat{\theta}$ and $\hat{\delta}$,
respectively. The constants $E_{C}$, $E_{J}$, $E_{J}^{R}$, $C$ and
$C_{g}$ are the single electron charging energy of the island
between the small junctions, the sum of the Josephson energy of the
two small junctions, the large junction Josephson energy, the total
capacitance shunting the large junction, and the gate capacitance,
respectively. Here $\varphi_{0}=\hbar/2e$ is the reduced flux
quantum. The control parameters $U(t)=U_{rf}\left( t\right)
\cos\Omega t$ and $I(t)=I_{rf}\left(  t\right)  \cos\omega t$ are
analogous to electromagnetic probe fields in an atomic system and
induce a charge excitation of the write port and a phase excitation
of the read port, respectively. This hamiltonian has been written
supposing that the offset gate charge and loop flux have been
compensated to operate at the optimal bias point where the charge
$\frac{\partial\hat{H}}{\partial U}$ and the flux $\frac{\partial
\hat{H}}{\partial I}$ have zero mean value in both the ground
$\left\vert 0\right\rangle $ and first excited $\left\vert
1\right\rangle $ states of $\hat{H}_{box}$. Under these conditions,
the qubit is minimally sensitive to charge and flux noise
\cite{quantronium}.

If we keep these two lowest states in the Hilbert space of the circuit we
obtain an effective qubit hamiltonian{}

\begin{widetext}
\begin{equation}
\hat{H}_{eff}=\frac{2C_{g}U\left( t\right) }{e}E_{C}\sigma _{X}-\frac{E_{J}}{%
2}\sigma _{Z}+\hbar \omega _{p}\left( 1+\lambda \sigma _{Z}\right)
a^{\dag }a-\mu \left( 1+\frac{\lambda}{4} \sigma _{Z}\right)\left(
a+a^{\dag }\right) ^{4}-f\left( a+a^{\dag }\right) I\left( t\right)
\tag{1}
\end{equation}
\end{widetext}

where
\begin{align*}
\omega_{p}  & =\sqrt{\frac{E_{J}^{R}}{\varphi_{0}^{2}C}}\\
\lambda & =\frac{E_{J}}{4E_{J}^{R}}\\
\mu & =\frac{E_{C}^{R}}{12}=\frac{1}{12}\frac{\left(  e\right)  ^{2}}{2C}\\
f  & =\varphi_{0}\left(\frac{2E_{C}^{R}}{E_{J}^{R}}\right)^{1/4}
\end{align*}

The photon annihilation operator $a$ is related to $\hat{\delta}$ by
\[
\hat{\delta}=\frac{a+a^{\dag}}{\left(  E_{J}^{R}/2E_{C}^{R}\right)
^{1/4}}\] which represents the decomposition of the gauge-invariant
phase difference into annihilation and creation operators of the
\textquotedblleft plasma\textquotedblright \thinspace mode whose
bare frequency is $\omega_{p}$. The operators $\sigma_{X}$ and
$\sigma_{Z}$ are the Pauli spin operators and $E_{C}^{R}$ is the
single electron charging energy of the readout junction. In this
effective hamiltonian, the expansion of $\cos\hat{\delta}$ is
carried out only to the first anharmonic term, which describes the
non-linear resonator dynamics with sufficient accuracy for a
bifurcation readout.

Let us describe the role of each term in (1). The first term
describes the influence on the qubit of the charge port drive\emph{
}which is used to manipulate its state. The second term is the
Larmor term $\omega _{01}=E_{J}/\hbar$. We have supposed here that
the ratio $E_{J}/E_{C}$ is sufficiently small that corrections to
the Larmor frequency involving $E_{C}$ are small. To model the
behavior of qubit samples with an appreciable $E_{J}/E_{C}$ ratio,
we would keep higher order terms, yielding renormalized values of
the coefficients in (1). The third term describes the dominant
coupling between the qubit and the resonator. Note that this term
commutes with the hamiltonian of the qubit when $U=0$, offering the
possibility of quantum non-demolition measurements. The fourth term
describes a decrease in the frequency of the resonator when its
photon population increases \cite{Siddiqi-plasma}. Finally, the
fifth term describes the excitation of the resonator by the drive
current applied through the phase port. When the drive current is
increased while its frequency is sufficiently below $\omega_{p}$ the
system becomes metastable with two possible dynamical states with
different oscillation amplitudes, i.e. two possible photon
populations \cite{Siddiqi-JBA-PRL}. We exploit this bistability for
our readout, which we describe in the next section.

\section{Qubit readout}

It is clear from the hamiltonian (1) above that the dynamics of the
non-linear resonator depend on the value $\sigma_{Z}=\pm1$
corresponding to the state of the qubit. In particular, the small
oscillation \textquotedblleft plasma\textquotedblright\ frequency
$\omega_{p}^{\rm eff}=\omega_{p}\left( 1\pm\lambda\right)  $ varies
with the qubit state. We probe the nonlinear resonator by sending
down the phase port transmission line a microwave pulse with carrier
frequency $\omega=\omega_{p}-\Delta\omega$, such that the detuning
$\Delta\omega>\frac{\sqrt{3}}{2Q}\omega_{p}$ where $Q$ is the
quality factor of the plasma resonance \cite{Dykman}. In our
circuit, the damping of the plasma resonance arises from the
characteristic transmission line impedance
$Z_{c}=50\,\mathrm{\Omega}$ and thus
$Q=Z_{c}C\omega_{p}\simeq10-20$. For this value of detuning, when
ramping up the drive current $I_{rf}$ the
resonator switches from one dynamical state to another when%

\[
I_{rf}>I_{B}\left(  \omega,\omega_{p}^{\rm eff}\right)
\]
where $I_{B}$ is the bifurcation current with expressions given in
\cite{MQC-Siddiqi}. Therefore, by choosing the maximum pulse amplitude%

\[
I_{B}\left[  \omega,\omega_{p}\left(  1-\lambda\right)  \right]  <I_{rf}%
^{\rm \max}<I_{B}\left[  \omega,\omega_{p}\left(  1+\lambda\right)
\right]
\]
we can determine, by measuring if the resonator has switched or not,
whether the qubit was in state $\left\vert 0\right\rangle $ or
$\left\vert 1\right\rangle $.

The dynamical states of the resonator differ in both the amplitude
and phase of the forced oscillations at frequency $\omega$. In this
work, we have chosen to use a reflectometry setup in which all the
information about the resonator state is carried by the reflected
drive signal phase $\phi$. This last property occurs because the
probed circuit is not intrinsically dissipative (in absence of
quasi-particles, which is very well realized in our measurements)
and the power reflected from the chip is equal to the incident power
in steady state.
\begin{table*}[t]
\caption{Parameters for two measured qubit samples. The readout
frequency was 1.55 GHz and 1.70 GHz for samples A and B,
respectively. The detuning was 6\% of $\omega_{p}$.
The parameter $\eta$ is the discrimination power of the readout.}%
\label{tab:table1}
\begin{ruledtabular}
\begin{tabular}{ccccccccc}
\\
Sample & $\omega _{01}/2\pi \mathrm{(GHz)}$ & $E_{J}/E_{C}$ &
$T_{1,\rm typical}\mathrm{(\mu s)}$ & $T_{2}\mathrm{(ns)}$ & $T_{\rm
echo}\mathrm{(ns)}$ & $\eta _{\rm \exp }$ & $\eta _{\rm calc}$ &
$\eta _{\rm \exp }/\eta_{\rm calc}$
\\
\hline
A & $9.513$ & $2.7$ & $4.0$ & $320$ & $400-500$ & $.48$ & $.70\pm.05$ & $.69$ \\
B & $18.989$ & $6.0$ & $1.0$ & $300$ & $300$ & $.61$ & $.70\pm.05$ & $.87$ \\
\end{tabular}
\end{ruledtabular}
\end{table*}
A further advantage of our non-linear resonator is that the
switching is strongly hysteretic. Once a switching event has
occurred we can decrease the drive current $I_{rf}$ to a value
which, while much smaller than $I_{B}\left[ \omega,\omega_{p}\left(
1-\lambda\right)  \right]  $, is still higher than the reverse
bifurcation \textquotedblleft retrapping\textquotedblright current
$I_{\bar{B}}$. This latching property conserves the information
about the qubit state acquired during a small time interval
$\tau_{m}$ in the resonator and allows us to probe the reflected
phase $\phi$ during a time typically longer than $\tau_{m}$.

In Fig. 2, we present a typical histogram of the reflected drive
signal phase $\phi$ corresponding to a drive current $I_{rf}$ which
causes the resonator to switch, on average, half of the time. The
histogram has 800,000 counts acquired in $200\thinspace\mathrm{ms}$.
For qubit measurements shown later, histograms with only 10,000 are
used. The shape of the readout pulse used is schematically shown in
the inset of Fig. 2. The rise time of the pulse is typically
$20-40\thinspace\mathrm{ns}$, the maximum current $I_{rf}^{\max}$ is
applied for $40-120\,\mathrm{ns}$ and the latched section lasts
$120\,\mathrm{ns}$, during which the recorded reflected signal phase
$\phi$ is bimodal, with values differing by about $124^{\circ}$. We
have chosen the phase reference so that the value $\phi=\phi_{\rm
low}=-62^{\circ}$ corresponds to the resonator in its initial state,
while $\phi=\phi_{\rm high}=62^{\circ}$ corresponds to the
resonator having switched. We define the switching probability $P_{\rm switch}%
\left(  I_{rf}^{\max},\Delta\omega,\left\langle \Psi\left\vert
\sigma _{Z}\right\vert \Psi\right\rangle \right)  $, where
$\left\vert \Psi \right\rangle $ is the state of the qubit, as the
weight of the histogram that lies above $\phi=\frac{\phi_{\rm
low}+\phi_{\rm high}}{2}=0$.

\begin{figure}[b]
\includegraphics[width=3.4in]{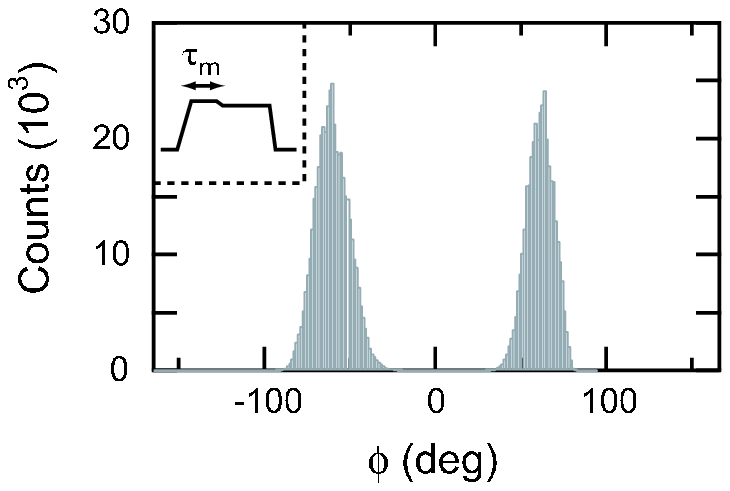}\caption{Typical histogram of the phase
of the reflected signal in the JBA readout when the maximum rf drive
current is chosen so that the resonator switches approximately half
of the time. The switching probability $P_{\text{switch}}$ is
defined as the fraction of the histogram lying above $\phi=0$. The
inset shows schematically the envelope of the readout pulse sent to
the phase port. The qubit influences the switching probability
during the
time interval $\tau_{m}$ which here was 40\,ns.}%
\label{FigSampleWL}%
\end{figure}

\section{Coherence results}

We now present experimental results on two different qubit samples
whose characteristic parameters are listed in Table I, along with a
summary of our results. In the figures that follow, we only show
data for sample A.

\bigskip

\begin{figure}[b]
\includegraphics[width=3.4in]{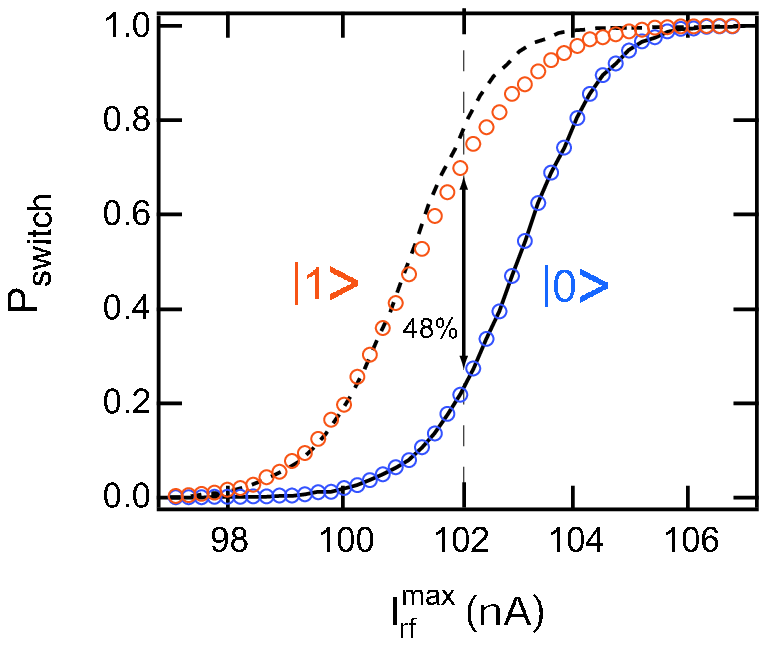}\caption{Switching probability as a
function of maximum drive current and qubit state for sample A.
Dotted line represents value of drive current at which maximal
discrimination power is observed. The width in current of the curves
is in reasonable agreement with numerical simulations (data not
shown). The solid line connects the observed data points in the
$\left\vert 0\right\rangle $ state and the dashed line is a copy of
the solid line horizontally shifted to overlap the $\left\vert
1\right\rangle$
state data at low values of $P_{switch}$.}\label{FigSampleWL}%
\end{figure}

We first characterized our readout by measuring $P_{\rm switch}$ as
a function $I_{rf}^{\max}$ and $\left\vert \Psi\right\rangle $, as
shown in Fig. 3. The blue circles correspond to the qubit in its
ground state, obtained by letting the qubit relax spontaneously,
while the red circles correspond to the qubit in its first excited
state obtained by applying a $\pi$ pulse, which will be discussed
below. An important remark is that only a slight change in shape of
$P_{\rm switch}\left(  I_{rf}^{\max}\right)  $ between the two qubit
states is observed, which indicates that the switching process
itself does not contribute strongly to the relaxation of the qubit.
In cases where the readout is suspected to induce significant
relaxation, the switching probability curve for the qubit excited
state displays a pronounced kink and can be obtained by a weighted
average of the observed curve for the ground state and the
prediction for the excited state \cite{Delft-Kink,Nakamura-recent}.
The discrimination power of
the qubit readout is defined as%

\[
\eta=\underset{I_{rf}^{\max}}{\max}\left[  P_{\rm switch}\left(
\left\langle \sigma_{Z}\right\rangle _{\Psi}=1\right) -P_{\rm
switch}\left(  \left\langle \sigma_{Z}\right\rangle
_{\Psi}=-1\right)  \right]
\] and its observed $(\eta_{\rm exp})$ and predicted $(\eta_{\rm calc})$ values
are given in Table I. Numerical simulations \cite{Vijay-theory} of
the full circuit have been used to compute the predicted values of
$\eta$. Note that several competing factors enter this calculation,
yielding similar values for samples A and B. The error bars reflect
uncertainties in the values of stray reactances on chip and the
precise resonator temperature.

The observed discrimination power is about $15-30\,\%$ smaller than
expected, and we attribute this loss to spurious on-chip defects. In
a set of experiments to be described in a later publication, we used
two readout pulses in succession to determine that a $15-30\,\%$
loss of qubit population occurs even before the resonator is
energized to its operating point. As photons are injected into the
resonator, the effective qubit frequency is lowered due to a Stark
shift via the phase port \cite{AC-Stark-Blais}. When the Stark
shifted frequency coincides with the frequency of an on-chip defect,
a relaxation of the qubit occurs. Typically, the qubit frequency
spans $200-300\,\mathrm{MHz}$ before the state of the qubit is
registered by the readout, and $3-4$ spurious resonances are
encountered in this range.

For future measurements, we have developed a new method to counter
this effect. When applying a readout pulse via the phase port, we
apply a compensating pulse via the charge port which Stark shifts
the qubit to higher frequencies. When balancing these pulses, we
have successfully reduced the net frequency shift to below
$20\,\mathrm{MHz}$ and have minimized population loss to defects
before the resonator switches. To increase the expected
discrimination power to unity, we must use samples with either a
larger qubit $E_{J}$ or a stronger phase coupling between the qubit
and readout resonator. The latter can be accomplished by using a
resonator with two Josephson junctions in series.

Having characterized our readout discrimination power, we performed
a series of experiments to assess the coherence of our qubit, namely
the measurements of $T_{1}$, $T_{2}$, $T_{\rm echo}$ and
$\widetilde{T}_{2}$. These times characterize the decay of the
excited state population after a $\pi$ pulse, the decay of Ramsey
fringes, the decay of the echo signal after a $(\pi, \pi/2, \pi)$
pulse sequence and the decay of the Rabi oscillations, respectively.

\begin{figure}[t]
\includegraphics[width=3.4in]{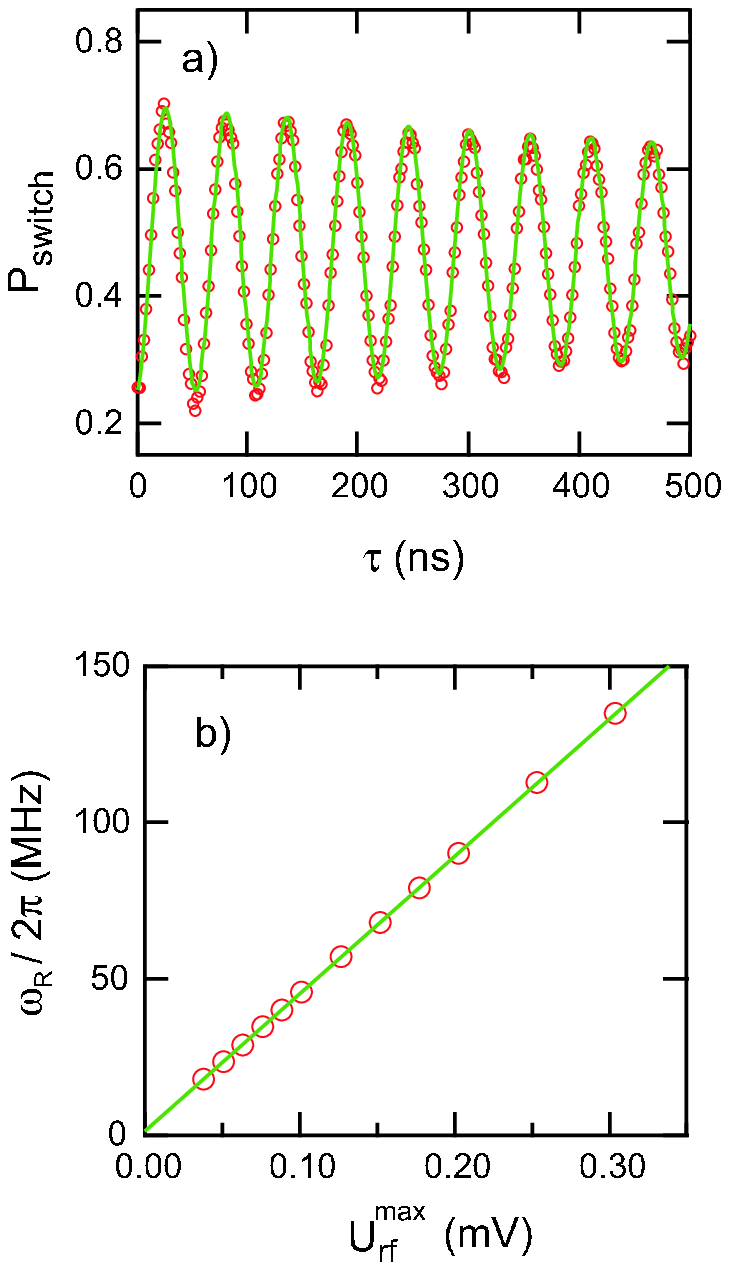}\caption{a) Rabi oscillations of the
switching probability of qubit sample A as a function of the
duration $\tau$ of a square pulse applied on the gate with maximum
amplitude $U_{rf}^{\max }=0.12\mathrm{mV}$. Solid green curve is an
exponentially decaying sinusoidal fit with
$\widetilde{T}_{2}=1.6\,\rm{\mu s}$. Total acquisition time is 3mn
and duty cycle is 16$\mu\mathrm{s}$, set by $T_{1}$ (see below). b)
Rabi oscillation frequency measured in (a) as a
function of $U_{rf}^{\max}$. Green line is expected linear dependence.}%
\label{FigSampleWL}%
\end{figure}

We first applied to the charge port a pulse at the Larmor frequency
$\omega_{01}$ of varying duration $\tau$ and amplitude $U_{rf}^{\rm
max}$, which performs a rotation of the qubit about $\sigma_{X}$,
followed by a readout pulse on the phase port. The resulting Rabi
oscillations in the switching probability signal are plotted in Fig.
4a for varying $\tau$ and fixed $U_{rf}^{\rm max}$. Near $\tau=0$ we
observe the $P_{\rm switch}$ corresponding to qubit being in the
$\left\vert 0\right\rangle $ state. As the pulse length increases,
$P_{\rm switch}$ increases, goes through a maximum where the qubit
is purely in the $\left\vert 1\right\rangle $ state, defining at
this point the length of a $\pi$ pulse. The switching probability
then decreases back to the $\left\vert 0\right\rangle $ state value,
indicating a full $2\pi$ rotation of the Bloch vector. This pattern
repeats itself but with diminishing contrast. The decay time
$\widetilde{T}_{2}$ is in the range $0.8-1.7\,\mathrm{\mu s}$
depending on the sample and precise biasing condition. The Rabi
oscillation frequency $\omega_{R}$ is plotted as a function of
$U_{rf}^{\rm max}$ in Fig. 4b. A linear dependence of $\omega_{R}$
with $U_{rf}^{\rm max}$ is observed, in agreement with theory. The
shortest $\pi$ pulse we generated was $2\,\mathrm{ns}$ long, and was
used in the echo experiments described below.

Having calibrated the $\pi$ pulse, we then performed a qubit energy
relaxation measurement by introducing a waiting time $t_{w}$ between
the $\pi$ pulse and the readout pulse. The decay of $P_{\rm switch}$
with $t_{w}$, shown in Fig. 5, is well fitted by a single
exponential, defining $T_{1}$. For sample A, $T_{1}$ was in the
range $2.5-5\,\mathrm{\mu s}$, and for sample B, $T_{1}$ was between
$1.0-1.3\,\mathrm{\mu s}$. The values of $T_{1}$ obtained with our
dispersive readout are comparable with the results of Vion
\textit{et al.} \cite{quantronium}, and are significantly shorter
than the values expected from coupling to a well thermalized
$50\,\mathrm{\Omega}$ microwave environment shunting the qubit. The
loss mechanisms giving rise to the observed energy relaxation are
not understood at this time.

\begin{figure}[t]
\includegraphics[width=3.4in]{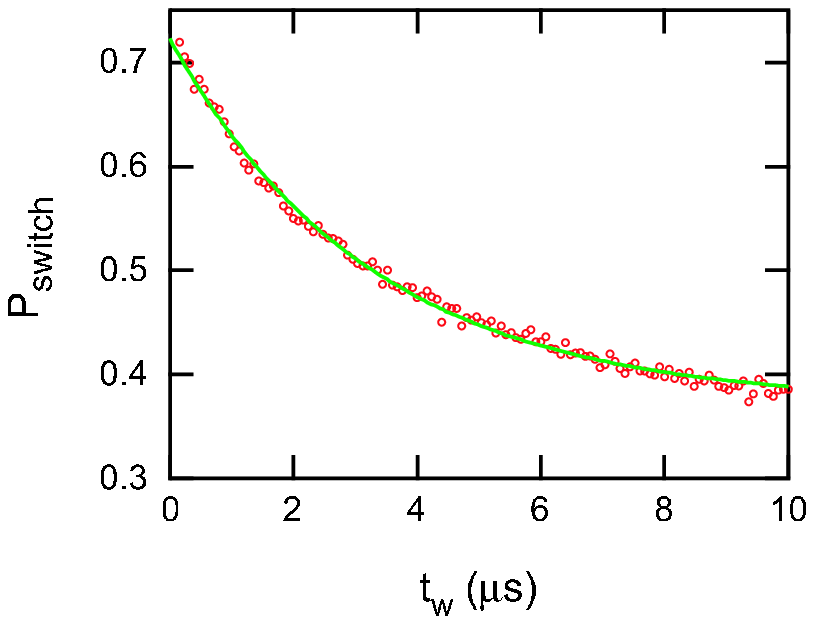}\caption{Decay of the excited state
switching probability after preparing the qubit in the excited state
by a $\pi$ pulse, as a function of the waiting time $t_{w}$ between
the preparation pulse and the readout pulse. Data for sample A.
Solid green curve is
an exponential fit with a 3.2$\mu\mathrm{s}$ decay constant.}%
\label{FigSampleWL}%
\end{figure}

Following measurements of the qubit energy relaxation, we performed
a Ramsey fringe experiment to determine the phase coherence of the
qubit. In this experiment, two $\pi/2$ pulses were applied to the
charge port of the qubit at a frequency $10-20\,\mathrm{MHz}$
detuned from $\omega_{01}$ followed by a readout pulse on the phase
port. A free evolution time $\Delta t$ was introduced between the
two $\pi/2$ pulses. In Fig. 6, $P_{\rm switch}$ is plotted as a
function of $\Delta t$. In the Ramsey sequence, the first $\pi/2$
pulse tips the Bloch vector from the north pole to the equatorial
plane. During the time $\Delta t$, the Bloch vector precesses around
the equatorial plane and is then rotated again by the second $\pi/2$
pulse. For $\Delta t=0$, the two $\pi/2$ pulses back to back act as
a single $\pi$ pulse and the observed value of $P_{\rm switch}$
corresponds to the qubit being in the $|1\rangle$ state. As $\Delta
t$ increases, $P_{\rm switch}$ decreases until it reaches the value
corresponding to the qubit being in the $|0\rangle$ state,
corresponding to a free evolution time $\Delta t$ in which the Bloch
vector makes a $\pi$ rotation in the equatorial plane. The switching
probability then continues to increase for larger values of $\Delta
t$ until it reaches a maximum value, corresponding to a time $\Delta
t$ where the Bloch vector makes a full $2\pi$ rotation in the
equatorial plane. This oscillatory pattern then repeats but with
decreasing contrast corresponding to the loss of phase coherence
with time. The Ramsey fringes decay in a time $T_{2}$ which has a
component due to energy relation and one due to pure dephasing:
$1/T_{2}=1/\left( 2T_{1}\right)  +1/T_{\varphi}$, where
$T_{\varphi}$ represents pure dephasing. In our measurements,
$T_{2}$ is dominated by pure dephasing. For sample A,
$T_{2}=320\,\mathrm{ns}$, and for sample B,
$T_{2}=300\,\mathrm{ns}$.

\begin{figure}[t]
\includegraphics[width=3.4in]{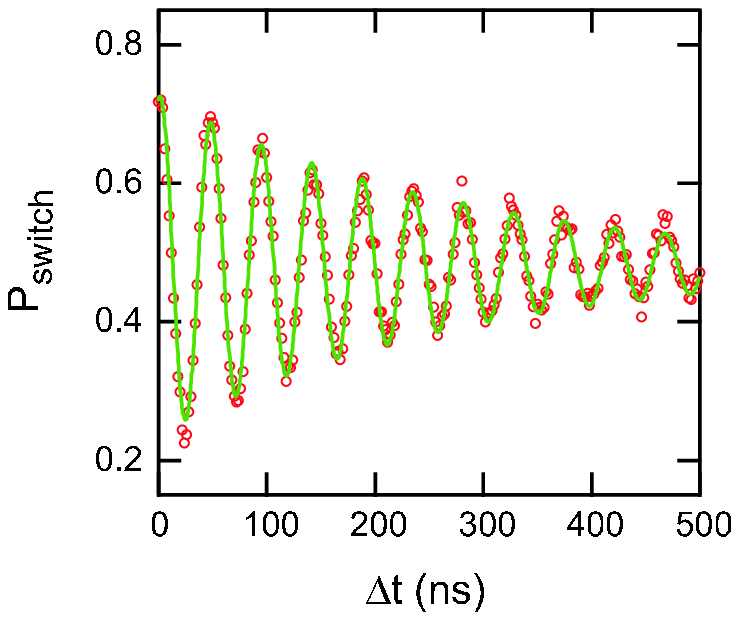}\caption{Ramsey fringes obtained with
two $\pi/2$ pulses separated by the time interval $\Delta t$. The pulse
frequency was detuned from the Larmor frequency by 20MHz. The green curve is a
exponentially decaying sinusoid fit. The decay time $T_{2}$ is 320ns. Same
acquisition conditions as in Fig. 4.}%
\label{FigSampleWL}%
\end{figure}

In order to correct dephasing of the qubit due to low frequency
noise \cite{Vion-echo, Delft-recent-echo}, we performed an echo
experiment in which we inserted a $\pi$ pulse in the middle of the
\begin{figure}[t]
\includegraphics[width=3.4in]{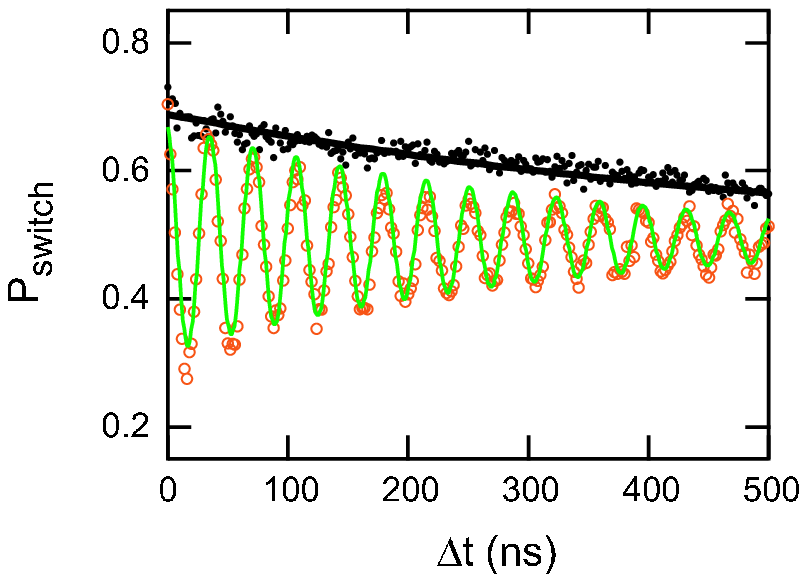}\caption{Result of an echo experiment
where a $\pi$ pulse was kept in the middle of the two $\pi/2$ pulses
separated by interval $\Delta t$ (black dots). The corresponding
Ramsey fringe data is shown with red circles. The thick
black curve is an exponentially decaying fit. }%
\label{FigSampleWL}%
\end{figure}
two $\pi/2$ pulses of the Ramsey sequence. A set of Ramsey fringes
and its corresponding echo decay are shown in Fig. 7 for sample A.
For this sample, the decay constant was increased to
$400-500\,\mathrm{ns}$ using the echo technique. For sample B, the
echo technique did not increase the phase coherence time. We believe
that for sample B, which has a large ratio of $E_{J}/E_{C}$ and is
protected from $1/f$ offset charge noise, the dominant source of
dephasing is due to broadband noise emanating from residual photons
in our readout resonator \cite{Delft-theory-dephasing}, thus
explaining the inefficacy of the echo sequence. It is possible that
the $50\,\mathrm{\Omega}$ environment shunting the qubit on the
phase port side was not fully thermalized to the refrigerator
temperature of $10\,\mathrm{mK}$. For sample A, where an improvement
was observed with the echo sequence, there are likely two
contributing factors. First, the ratio $E_{J}/E_{C}$ is much smaller
and offset charge noise played a stronger role. The low frequency
component of this noise can be corrected using an echo sequence.
Second, we added more cryogenic attenuation in the transmission
lines directly coupling to the phase port to reduce the resonator
temperature, thereby potentially reducing the number of excess
photons in the readout resonator and their associated dephasing.

\section{Conclusion}

In conclusion, we have successfully implemented a non-linear
dispersive readout of the quantronium qubit using the Josephson
Bifurcation Amplifier. The readout speed and discrimination power
show a significant improvement when compared with the DC switching
readout used in the original quantronium measurements
\cite{quantronium}. Perhaps even more important, in the present
readout scheme, the total\emph{ }measurement time is much smaller
than $T_{1}$, and it is possible to carry out experiments with
multiple readout pulses to determine the information flow during a
qubit readout and to account for any losses in qubit population.
This important aspect can be used to determine the degree to which
the measurement is quantum non-demolishing, and will be described in
later publications.

The authors are grateful to Patrice Bertet, Daniel Esteve, Steve
Girvin, Daniel Prober, and Denis Vion for helpful discussions. This
work was supported by ARDA (ARO Grant No. DAAD19-02-1-0044), the
Keck foundation and the NSF (Grant No. DMR-0325580).

\end{document}